\documentclass[namedreferences,hyperref,optionalrh]{spr-sola}
\usepackage{graphicx}        
\usepackage{color}           

\usepackage{amsmath}

\chardef\us=`\_

\begin{document}

\begin{frontmatter}
\title{The SynCOM Flow Tracking Challenge}

\author[addressref={aff1,aff2},email={moraesfilho@cua.edu}]{\inits{V.M.F.}\fnm{Valmir}~\snm{Moraes Filho}\orcid{0000-0002-5447-9964}}
\author[addressref={aff1,aff2},email={uritsky@cua.edu}]{\inits{V.M.U.}\fnm{Vadim}~\snm{Uritsky}\orcid{987-654-3210}}
\author[addressref={aff2}]{\inits{B.J.T.}\fnm{Barbara}~\snm{Thompson}}

\address[id=aff1]{Department of Physics, Catholic University of America \\
620 Michigan Ave NE \\
Washington, DC 20064, USA}
\address[id=aff2]{NASA Goddard Space Flight Center, Greenbelt, MD, USA}

\runningauthor{Moraes Filho et al.}
\runningtitle{\textit{Solar Physics} Example Article}

\begin{abstract}
Understanding solar wind flows is crucial for unraveling the dynamics of the Sun's corona and improving space weather forecasting. However, the complex nature of the solar wind and the absence of reliable ground-truth data present significant challenges to current tracking methods. To address this, the Flow Tracking Challenge was established, providing a platform for testing and refining flow tracking algorithms using synthetic images generated by the Synthetic Coronal Outflow Model (SynCOM). The challenge is divided into two phases: a preliminary phase focusing on simpler flow scenarios and a main phase featuring more complex synthetic images that mimic solar outflows. These phases allow researchers to evaluate their methods under various simulated conditions. SynCOM's synthetic data benchmark improves accuracy in velocity estimations. This paper presents the preliminary results of the Flow Tracking Challenge, where the synthetic images from SynCOM are used to test solar wind velocity tracking methods. Initial findings demonstrate SynCOM's value as a benchmark, guiding improvements for upcoming missions like PUNCH.

\end{abstract}
\end{frontmatter}

\section{Introduction}  
\label{S-introduction}

Solar activity primarily drives space weather, which presents considerable threats to Earth's technological systems, such as satellites, power grids, and communication networks \citep{Knipp_2019}. Predicting space weather is crucial for mitigating the potential damage caused by solar events such as solar flares, coronal mass ejections (CMEs), and high-speed solar wind streams, which can lead to geomagnetic storms and other disruptive phenomena. The solar wind is a stream of magnetized plasma transporting energy from the Sun to Earth's magnetosphere. Recent studies have shown that the solar wind is highly structured and complex , featuring transient events such as coronal jets, plumes, and type 2 spicules \citep{Viall_2010,Rouillard_2011,Cranmer_2017,Richardson_2018}. Understanding these dynamic processes is essential for improving space weather predictions and mitigating their effects on Earth's technological systems \citep{Beedle_2022}.

The upcoming PUNCH (Polarimeter to Unify the Corona and Heliosphere) mission aims to significantly enhance our understanding of the solar wind. The high-resolution data it collects will enable three-dimensional assessments of solar wind formations ranging from the corona to the inner heliosphere \citep{DeForest_2022}. Researchers will benefit from the mission's ability to observe the development of these features with unparalleled detail. These observations will facilitate the mapping of solar wind velocity across different regions, providing critical insights into the mechanisms of solar wind acceleration and the formation of transient features such as plasma blobs \citep{Viall_2023}.

Various flow tracking algorithms have been developed to assess velocity in different areas of solar physics. Techniques like local correlation tracking (LCT) are used to detect displacements in photospheric features \citep{november_1988}, while balltracking identifies intensity minima within granules \citep{potts_2004,Attie_2009}, and its enhanced version, magnetic balltracking, incorporates magnetic field data for more precise tracking \citep{attie_2015}. Meanwhile, Distance-Time (DT) plotting is employed to visualize motion in CMEs \citep{Sheeley_1999}. DT plotting can also be used with the surfing transform, is used to track nearly parallel flows, smoothing out random fluctuations while retaining temporal resolution \citep{uristky_2009}; and also with the time-domain correlation method \citep{moraes_filho_2024}. Moreover, combining DT plotting with the Fourier transform creates 'speed spectra' for coronal regions, measuring velocities regardless of position, providing a distinct approach from purely feature-based techniques \citep{deforest_2014}. 

A significant challenge with existing flow tracking algorithms is the lack of validation and ground-truth data, which hinders the ability to accurately assess their performance. This gap makes it difficult to verify and fine-tune these methods, leaving uncertainties about which techniques are the most reliable under varying solar conditions. However, these algorithms are essential for missions like PUNCH, as they play a crucial role in mapping the solar wind and understanding its acceleration processes.

To address this issue, SynCOM (Synthetic Corona Outflow Model) was developed to provide a high-resolution, data-driven simulation of solar wind flows \citep{moraes_filho_2024}. SynCOM provides a reliable ground truth standard for testing and validating flow tracking algorithms by producing synthetic images that replicate realistic plasma fluctuations and noise \citep{moraes_filho_2024_zenodo}. It has already been successfully used to test these techniques in a controlled environment, helping bridge the accuracy and reliability gap between different flow tracking methods.

This paper presents the SynCOM flow tracking challenge structure. Section \ref{s-motivation_FT} covers the motivation behind the Flow Tracking Challenge and the need for accurate solar wind velocity measurements. Section \ref{S-methodology_FT} details the methodology, focusing on the exercises used to test the tracking algorithms. Section \ref{S-challenge-FT} discusses the specific challenges and limitations of current tracking methods. Finally, Section \ref{S-reults-FT} presents the results of the challenge, emphasizing SynCOM's role in validating tracking methods.

\section{Why this Challenge was created?}
\label{s-motivation_FT}

The Flow Tracking Challenge was motivated by the challenges we encountered while attempting to accurately map coronal outflows. During the initial stages of this research, we found significant uncertainties in the validation of velocity measurements in different flow tracking methods \citep{moraes_filho_2020}. The main issue was the absence of a reliable ground-truth standard, which led to inconsistencies in the results. To address this issue, we created SynCOM to assess and enhance these techniques. The IDL source code version of the SynCOM software suite is accessible at \url{https://doi.org/10.5281/zenodo.13357546} as detailed in \citep{moraes_filho_2024_zenodo}.

In June 2022, the Flow Tracking Challenge was formally introduced at the SHINE Workshop \citep{Moraes_Filho_2022}. Researchers were invited to apply their methods to synthetic Distance-Time (DT) plots, providing a benchmark for comparing performance, particularly relevant for the PUNCH mission, which was evaluating flow tracking methods for its data products.

In June 2022, the Flow Tracking Challenge was introduced at the SHINE Workshop, inviting researchers to test their methods on synthetic DT plots \citep{Moraes_Filho_2022}. The model offers a controlled benchmark for the upcoming PUNCH mission to evaluate effective flow tracking techniques, establishing a standard for comparison.

Based on feedback and ongoing research, we expanded the challenge by incorporating synthetic images that better simulate real solar wind observations, similar to those from STEREO observations.The increased complexity provides a more realistic setting for testing, allowing participants to more effectively adapt and cross-validate their techniques across a wider variety of solar wind conditions. The challenge now serves to validate both new and existing flow tracking algorithms, providing insight into their strengths and limitations. Ultimately, this initiative supports the development of more accurate methods for interpreting high-resolution solar wind data from missions like PUNCH.

\section{How the Challenge Works?}
\label{S-methodology_FT}

The Flow Tracking Challenge is divided into two primary stages: the preliminary challenge and the main challenge, each designed to assess different aspects of flow tracking algorithms. Hosted on the Catholic University of America website \citep{Moraes_Uritsky_2022}, the challenge provides synthetic datasets in multiple formats (FITS, SAV, and MOV), allowing participants to test their methods. Conducted as a blind test, participants are not given specific details about the simulated flows, replicating real-world conditions to ensure a comprehensive evaluation of their algorithm's performance.

The Flow Tracking Challenge is structured into two main stages: the preliminary challenge and the main challenge, each designed to test different aspects of flow tracking algorithms. Hosted on the Catholic University of America website \citep{Moraes_Uritsky_2022}, the challenge provides synthetic sets in multiple formats (FITS, SAV, and MOV), allowing participants to test their methods and submit their results for feedback. Conducted as a blind test, specific details about the simulated flows are not disclosed to the participants. This replicates real-world conditions, ensuring a thorough evaluation of algorithm performance.

\subsection{Preliminary Challenge}
\begin{figure}[ht]
    \centering
    \includegraphics[width=\linewidth]{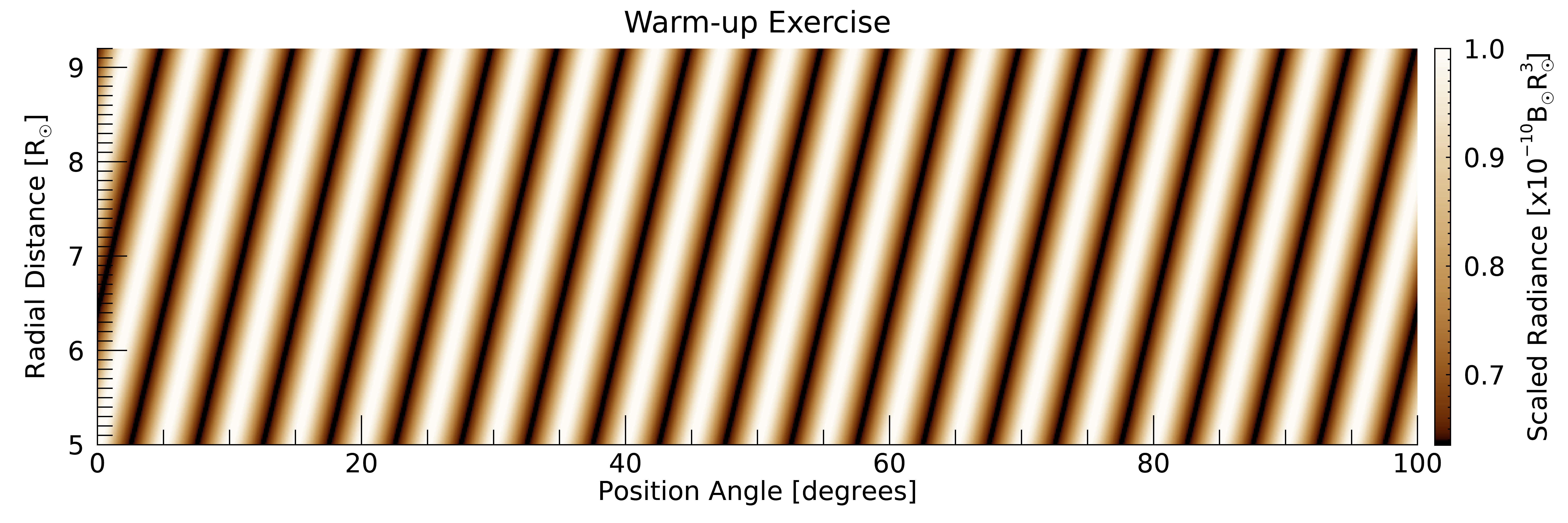}
    \caption{This model exemplifies a steady radial flow marked by consistently spaced, parallel features. Intended as an introductory exercise, it offers participants the opportunity to fine-tune their feature-tracking algorithms within a controlled setting. The predetermined flow velocity of 75 km/s and a period of 1200 seconds serve as reference metrics, allowing participants to calibrate their algorithms and ensure that they are equipped to tackle the challenge.}
    \label{fig:flow_tracking_warm_up}
\end{figure}

The preliminary phase is a controlled environment in which participants familiarize themselves with flow tracking using simpler synthetic data. This stage includes several models, each with varying temporal and spatial resolutions. The warm-up model (Figure \ref{fig:flow_tracking_warm_up}) features steady radial flow with a known velocity of 75 km/s and a period of 1200 seconds, allowing participants to calibrate their algorithms in a straightforward setting. The model's temporal resolution is 24 seconds and its spatial resolution is 435 km, providing reference metrics for participants to fine-tune their methods.

Models 1 and 2 build on this by introducing small-scale uniform flows with a finer temporal resolution of 12 seconds and the same spatial resolution. More advanced scenarios, such as Models 3 to 6, incorporate periodic disturbances and turbulence, using temporal and spatial resolutions similar to those of the STEREO-A/COR2 instrument (300 seconds and 9,744 km). These varied models allow participants to test their algorithms under a range of conditions, from simple, uniform flows to more complex, large-scale disturbances. In these scenarios, participants estimate the velocity by tracking the displacement of features across consecutive frames, using temporal and spatial resolutions to convert the pixel displacement into physical velocity measurements.

\begin{figure}
    \centering
    \includegraphics[width=\linewidth]{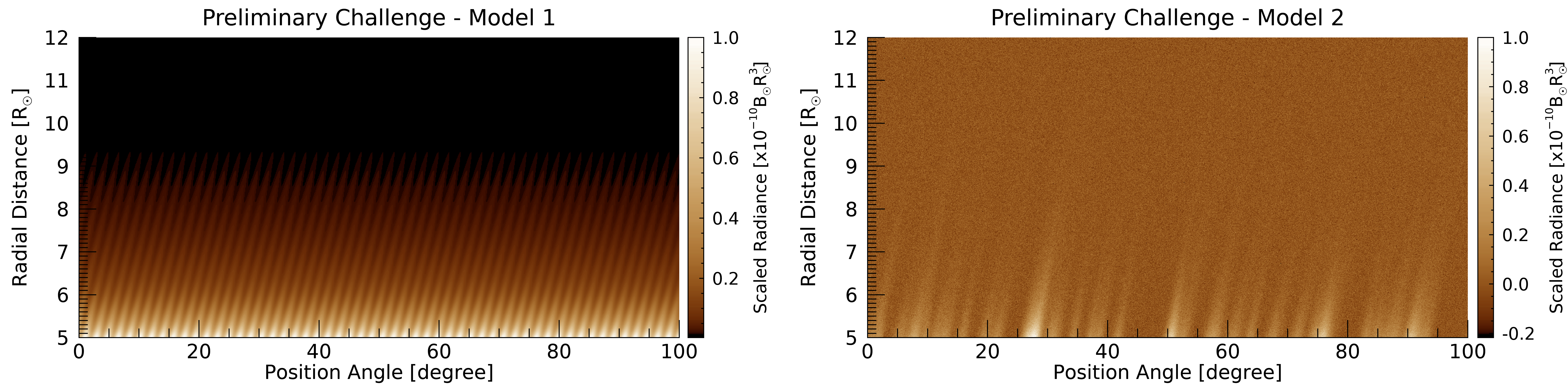} \\
    \includegraphics[width=\linewidth]{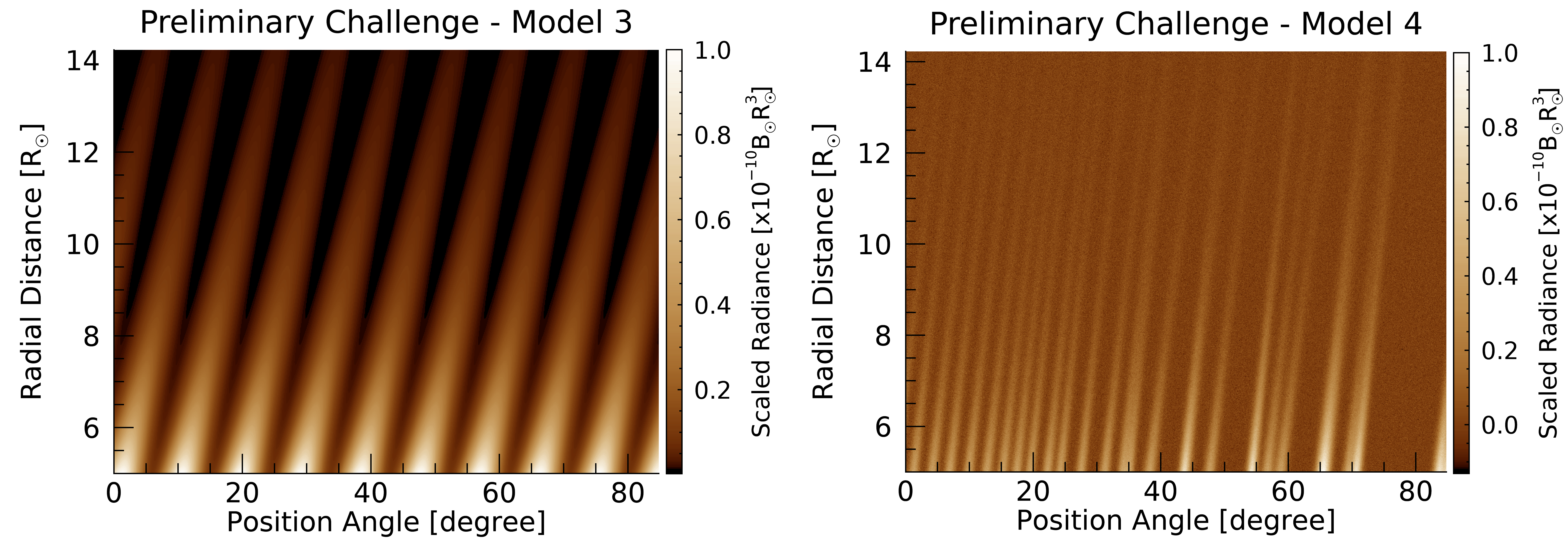} \\
    \includegraphics[width=\linewidth]{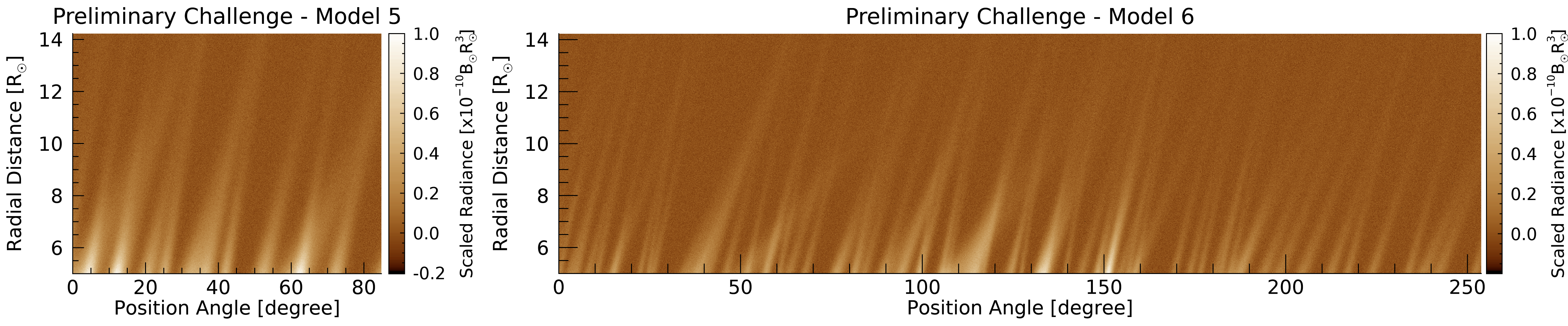}
    \caption{The images show six distinct models from the Preliminary Challenge, designed to test flow-tracking algorithms under varying conditions. Model 1 features a high temporal (12s) and spatial (435 km) resolution, with uniform small-scale radial flows. Model 2 maintains the same resolution, but introduces minor flow variations. Model 3 increases complexity, simulating large-scale periodic disturbances with lower temporal (300s) and spatial (9,744 km) resolution. Models 4, 5, and 6 provide increasingly irregular and turbulent flows, further challenging the participants' algorithms with a resolution similar to Model 3, which matches the settings used by STEREO-A/COR2. These diverse scenarios allow participants to test their velocity estimation methods in a range of solar wind conditions.}
    \label{fig:Flow_tracking_simple}
\end{figure}

Participants are encouraged to use any method they prefer to estimate the velocity of the flow present in the snapshot, as long as they adhere to the provided spatial (dr) and temporal (dt) resolutions. They achieve velocity estimation by tracking the feature displacement across consecutive frames, using the temporal resolution (dt) to determine the time interval between images and the spatial resolution (dr) to convert the pixel displacement into physical distance (km). The velocity (in km/s) is then calculated using the equation: 
 \begin{equation}
    Velocity = \frac{\text{Displacement (km)}}{\text{Time interval (seconds)}} = \frac{dr\text{ (km/pixel)}\times(\text{pixel displacement})}{dt\text{ (seconds)}} 
\end{equation}

This approach enables participants to generate velocity estimates and compare their results against the known velocity profiles of synthetic models, facilitating the validation and calibration of their flow tracking algorithms.

\begin{figure}
    \centering
    \includegraphics[width=\linewidth]{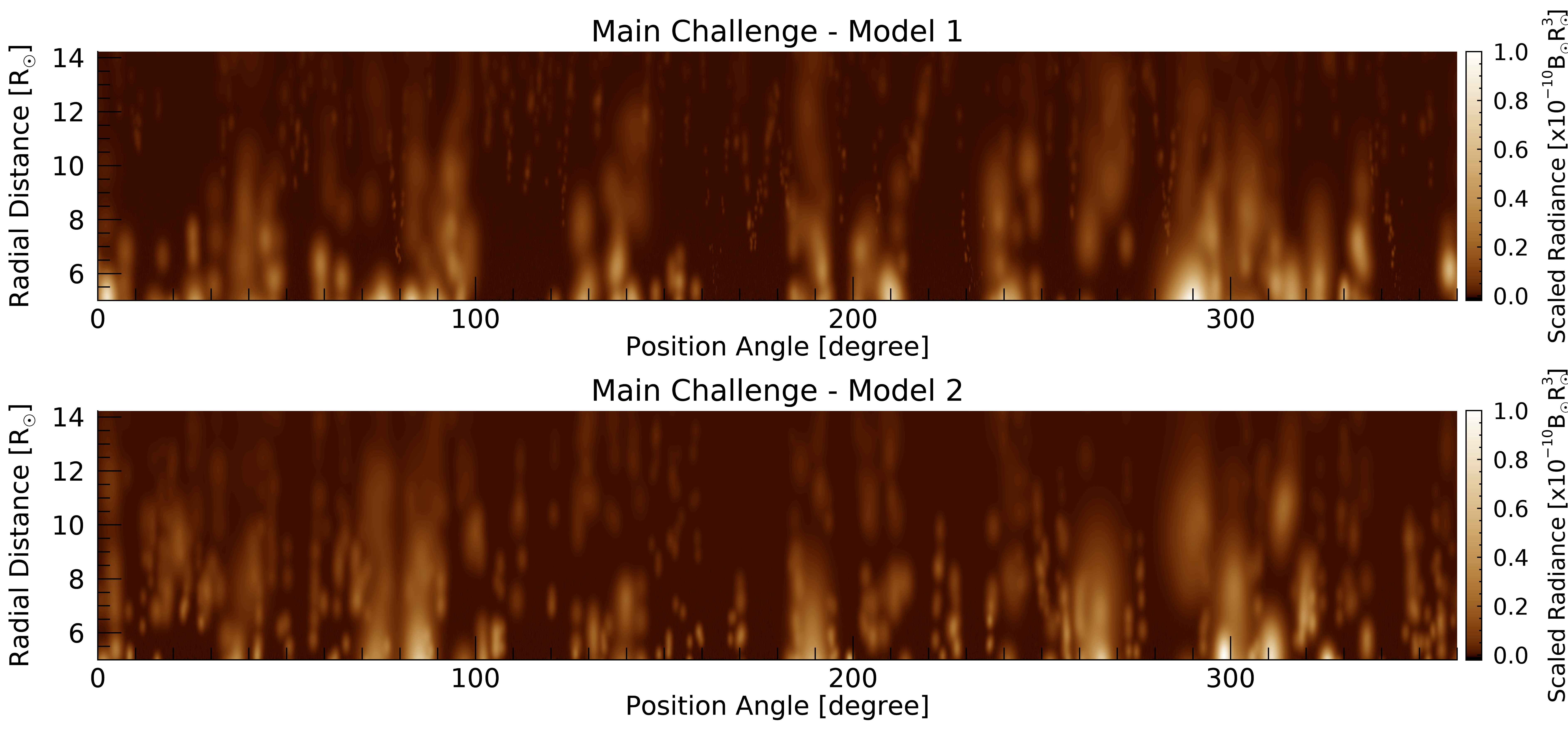}
    \caption{These images represent the 2023 version of SynCOM, included in the Main Flow Tracking Challenge for their flow-like dynamics. Using a time interval of 300 seconds and a radial distance increment of 9,744 kilometers, these models extend over 70.8 hours (850 frames) through radial distances ranging from 4 to 15 solar radii ($R_\odot$), encompassing position angles from 0 to 360 degrees. This matches the observational settings of the STEREO-A/COR2 instrument, further emphasizing the connection between synthetic data and real-world solar wind observations. These flow scenarios encourage a broader variety of flow trackers to test SynCOM, allowing participants to estimate velocities by tracking feature displacement in complex coronal conditions.}
    \label{fig:Flow_tracking_main}
\end{figure}
\subsection{Main Challenge}

The main challenge builds on the preliminary phase by presenting more complex simulations, including steady solar wind, turbulent regions, and CMEs. Unlike the preliminary stage, which focused primarily on DT plots, the main challenge provides synthetic images that closely mimic real solar observations. These images were generated using SynCOM software that produced realistic representations of solar wind outflows \citep{moraes_filho_2024_zenodo}. This phase encourages participants to use a wider range of tracking methods, including optical feature tracking techniques such as LCT and optical flow methods.

This phase features data with temporal and spatial resolutions similar to those of STEREO-A/COR2, further connecting the synthetic models to real observational settings. By applying these advanced techniques, participants can test and refine their algorithms, preparing them for future missions such as PUNCH.
\begin{figure}
    \hspace{0.0 cm} (a)  \\
    \includegraphics[width=\linewidth]{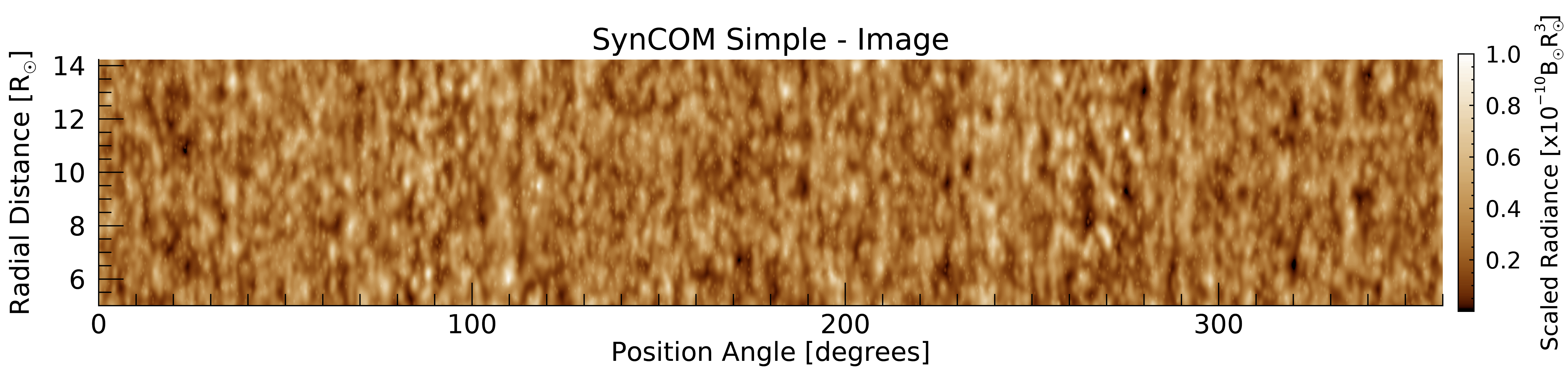} 
    \hspace{0.0 cm}(b) \\
    \includegraphics[width=\linewidth]{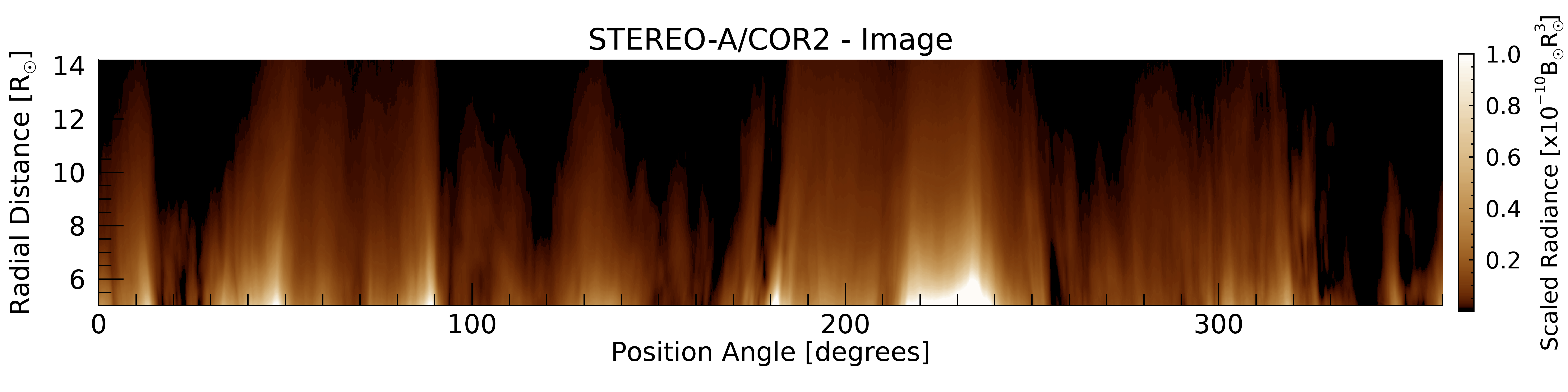}
    \hspace{0.0 cm} (c) \\
    \includegraphics[width=\linewidth]{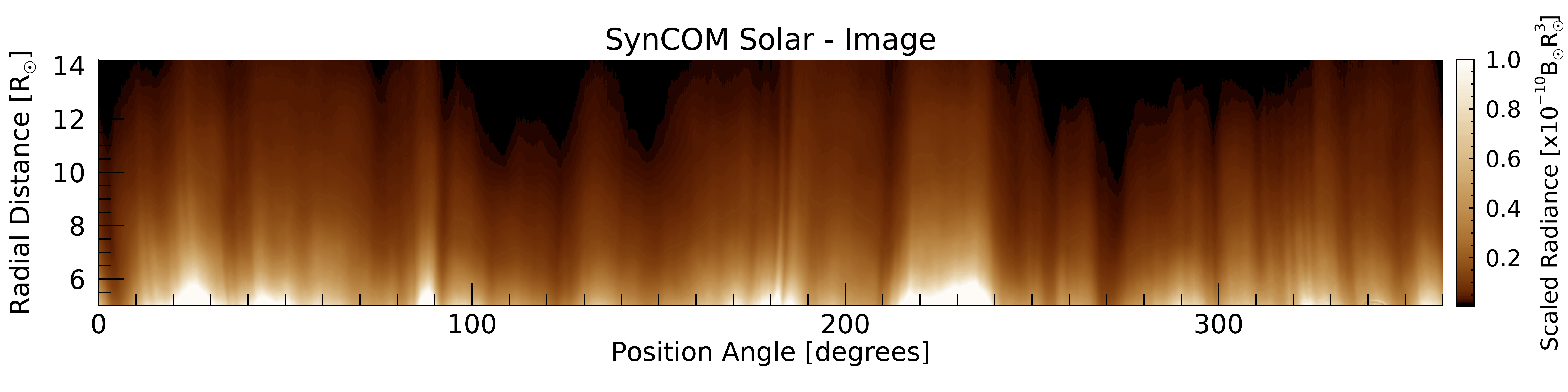}
    \hspace{0.0 cm} (d)\\
    \includegraphics[width=\linewidth]{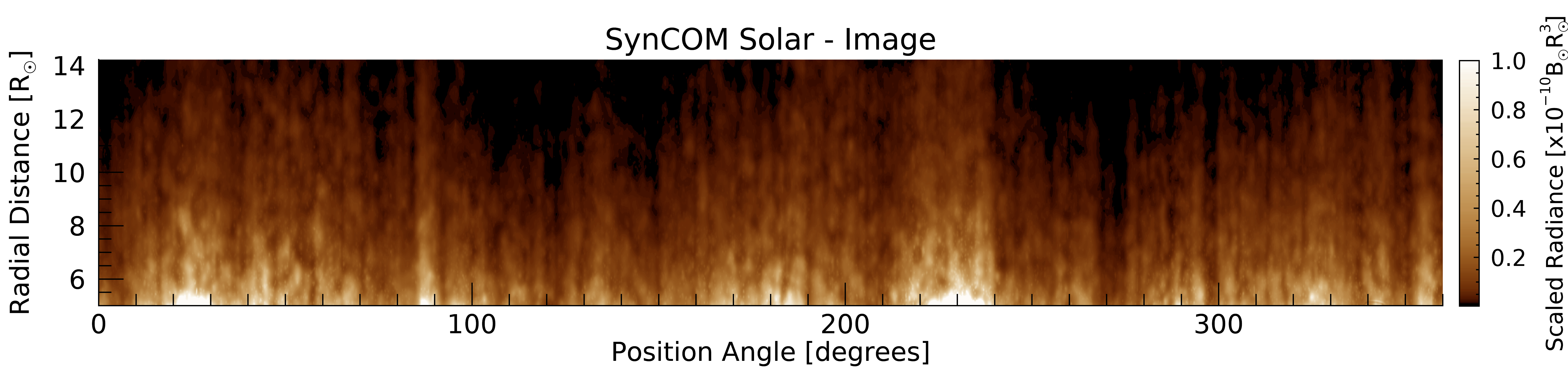}
    \caption{Comparison of a COR2 image with 2024 version of SynCOM. (a) SynCOM simulation with 5,000 blobs, serving as a warm-up example to tune up flow tracking methods. The velocity profile follows a sinusoidal profile with speeds between 150 and 350 km/s. (b)Training data from STEREO-A/COR2 showing solar outflows. (c) SynCOM simulation with larger blob sizes to match the intensity profile of COR2 data, capturing the overall structure but lacking fine details. (d) SynCOM simulation with smaller blobs, highlighting fine-scale structures, useful for studying small-scale outflows like jetlets, crucial in understanding solar dynamics. Source: \cite{moraes_filho_2024}}
    \label{fig:Flow_tracking_main_2}
\end{figure}

There are two versions of this stage, the first launched in 2023 (Figure \ref{fig:Flow_tracking_main}) and the second in 2024 (Figure \ref{fig:Flow_tracking_main_2}). The 2024 version introduces synthetic images specifically designed to analyze flow tracking methods with SynCOM \citep{moraes_filho_2024}. This version also features a warm-up simulation, offering participants a preliminary exercise to fine-tune their algorithms before tackling more complex scenarios. This version was developed as a more flexible framework that supports user-friendly customization, allowing participants to design their simulations with parameters tailored to meet their specific research requirements.

\section{Challenges}
\label{S-challenge-FT}

One of the main challenges of the Flow Tracking Challenge lies in its blind-test nature, where participants are provided with synthetic data without prior knowledge of the flow parameters. This setup mirrors real-world conditions, where accurate flow tracking is often hindered by image noise and background complexity in solar wind observations. Solar images typically contain noise due to instrumental limitations, cosmic interference, and variable solar activity, which complicates the detection of subtle solar wind features. Moreover, removing background elements, such as coronal structures and streamers, adds further complexity. Certain algorithms may perform better on background removal than others, affecting their overall accuracy.

The synthetic data sets used in the Flow Tracking Challenge are carefully designed to evaluate how well each algorithm handles different noise levels and intricate background conditions. By introducing controlled variations in these datasets, the challenge provides a reliable way to benchmark the robustness and adaptability of various flow tracking methods.

\section{Results and Expected Outcomes}
\label{S-reults-FT}

The preliminary results of the challenge have been encouraging. Many submissions successfully estimated velocity values within their standard deviation ranges, indicating the robustness of certain algorithms under specific simulated conditions. This iterative feedback loop has proven valuable, not only validating the use of synthetic models but also helping researchers refine their flow tracking techniques based on concrete performance data.

For the main challenge, which uses synthetic image data resembling real solar observations, the Time-Distance Correlation (TDC) method was applied. The initial results showed that TDC produced velocity estimates within acceptable standard deviation values, demonstrating its effectiveness in tracking flows under complex coronal conditions. However, further feedback highlighted the need for additional testing using other techniques, such as optical feature tracking and magnetic balltracking, to better understand the strengths and limitations of both algorithms and synthetic models. This continuous evaluation and comparison process will ultimately establish best practices for solar wind tracking, enhance model accuracy, and improve data interpretation for future missions like PUNCH.

\begin{acknowledgments}
The authors thank the Southwest Institute for hosting the STEREO-A/COR2 data prepared by Craig DeForest. We are grateful to the PUNCH team and the Heliophysics community for their support and participation in the SynCOM Flow Tracking Challenge. V.P.M.F. has been supported through the Partnership for Heliophysics and Space Environment Research (NASA grant No. 80NSSC21M0180). V.P.M.F. extends his gratitude to the NSF NCAR's Advanced Study Program's Graduate Visitor Program (GVP) for his fellowship in 2023.
\end{acknowledgments}

\bibliographystyle{spr-mp-sola}
\bibliography{sola_bibliography_example}

\end{document}